# Gamma-ray spectra of hexane in gas phase and liquid phase

Xiaoguang Ma[†*] and Feng Wang[*]

*eChemistry Laboratory, Faculty of Life and Social Sciences, Swinburne University of*

*Technology, PO Box 218, Hawthorn, Victoria 3122, Australia*

Theoretical gamma-ray spectra of molecule hexane have been calculated and compared with the experimental results in both gas (Surko *et al*, 1997) and liquid (Kerr *et al*, 1965) phases. The present study reveals that in gas phase not all valence electrons of hexane exhibit the same probability to annihilate a positron. Only the positrophilic electrons in the valence space dominate the gamma-ray spectra, which are in good agreement with the gas phase measurement. When hexane is confined in liquid phase, however, the intermolecular interactions ultimately eliminate the free molecular orientation and selectivity for the positrophilic electrons in the gas phase. As a result, the gamma-ray spectra of hexane become an "averaged" contribution from all valence electrons, which is again in agreement with liquid phase measurement. The roles of the positrophilic electrons in annihilation process for gas and liquid phases of hexane have been recognized for the first time in the present study.

PACS numbers: 34.80.-i, 36.10.-k, 78.70.Bj

Significant progress has been made to a more detailed understanding of the gamma-ray spectra of atoms and molecules [1-14], although more theoretical achievements have been obtained for atomic systems [6] than for molecules [3-5]. It is well known for both atoms and molecules that the core electrons only play a minor role in the electron-positron annihilation processes [1-5]. The core electrons are embedded respectively in the cores of all the atoms (centers) in a molecule and the nuclear repulsive potential prevents the positron from approaching these core electrons. As a result, the valence electrons dominate the gamma-ray spectra, which is analogous to the electron scattering processes of molecules.

---

[†] Permanent address: *School of Physics and Optoelectronic Engineering, Ludong University, Yantai, Shandong 264025, PR China.*

The interaction between a positron and a molecule or an atom is affected significantly by the total electrostatic potential (ESP) induced by all nuclei and electrons. For an isolated atom, i.e., a one-centre system, the total ESP has a spherical symmetry. As a result, the incoming positron will "feel" the same interaction at the same radius from the nuclei in space. For a multi-centre molecule, however, the total ESP usually does not exhibit a spherically symmetry except for highly symmetric cases such as methane ($CH_4$) and fullerene ($C_{60}$). The ESP of a molecule exhibits polarity with the electrophilic and nucleophilic (or, positrophilic) sites which do not exist in the case of atoms. It is known that not all valence electrons have the same contributions to the electronic processes of molecules [1-5], such as reactivity and ionization, depending on their orbital shape, symmetry and location. For example, in electron spectroscopy, electron transitions of a molecule can only happen for certain valence electrons (or orbitals), whereas other valence electron transitions of the same molecule are forbidden.

When a positron is involved, however, would each of the molecular electrons annihilate the positron equally, regardless their orbital energy, shape, symmetry and location? If yes, why could the positron annihilate core electrons differently from the valence electrons with a low probability? If no, what are the electrons which are most likely to be annihilated with the positron, i.e., what are the positrophilic electrons and how to identify the positrophilic electrons? In previous studies [3-5, 13], the authors indicated that not all valence electrons exhibit the same contributions to the positron-electron annihilation of atoms and small molecules in gas phase. In the present study, we provide more evidences on a "linear" polyatomic molecule of n-hexane ($C_6H_{14}$), to support the positrophilic electron model in annihilation of gamma-ray spectra of a molecule.

The efforts to study annihilation of positron and electron in polyatomic molecules, such as normal hexane [9, 12] (n-hexane), began nearly half a century ago. In the well-known measurement, the momentum distributions of normal hexane (n-$C_6H_{14}$) in the two photon annihilating positron-electron process were studied by Chuang and Hogg in 1967 [9] who developed a method based on analytic self-consistent field (SCF) wavefunctions of the carbon and hydrogen atoms of hexane. It was claimed that the positrons annihilate almost exclusively all the valence electrons in the C-H and C-C bonds of hexane [9]. The concept of averaged contribution of all valence electrons to the positron-electron annihilation of hexane was quickly accepted, as the theoretical results of Chuang and Hogg [9] (in gas phase) agreed well with their experimental measurement of hexane of Kerr, Chuang and Hogg [12] in liquid phase. Superposition of the multi-centred distribution of momenta approximately takes into account of interactions between the hexane molecules in liquid, hereby eliminating the valence electron orientation and selectivity which are insignificant in liquids. As a result, the total theoretical momentum distribution agreed with the measurement. The study marked a significant achievement at that time due to limited resources for more detailed studies of larger molecules.

Nobody reproduces the theoretical results of Chuang and Hogg for n-hexane [9] so far. Recently, Surko *et. al.* measured the gamma-ray spectra of a series of alkanes including hexane in low-pressure gas phase using the state-of-the-art high resolution Angular Correlation of Annihilation Radiation (ACAR) technique [2]. The gas phase measurement of hexane provides an excellent opportunity for a more detailed theoretical study of the low energy Doppler-shift of gamma-ray spectra of hexane as the intermolecular interactions among hexane molecules in liquid can be neglected in gas phase. It is noted that in the gas phase measurement, the Doppler-shift of the hexane molecule is given by 2.25 keV [2], whereas by 2.93 keV in the earlier liquid hexane measurement [12]. There is a significant difference of 0.68 keV in Doppler-shift of hexane between the two measurements, as the largest difference in Doppler-shift between the gas phase measurements of methane ($CH_4$) and dodecane ($C_{12}H_{26}$) is only 0.23 keV [2], which is nearly three times smaller than 0.68 eV. Such a significant discrepancy in Doppler-shift between two measurements cannot be simply explained as an instrumental/technical issue due to the advancement of the modern technology of Surko et al [2], although the earlier measurement [12] was supported by their theoretical study [9]. Such the large discrepancy in hexane measurements implies that a considerably large impact on the measurements attribute to the phase factor (i.e., gas phase and liquid phase). That is, different roles of the electrons in the molecule play in the annihilation process for gas and liquid conditions. The present study is to reveal the roles of electrons of n-hexane in the gamma-ray spectra quantum mechanically in gas phase and liquid phase.

The molecular electronic wave-functions of n-hexane are calculated using the Gaussian09 computational chemistry package [15]. The model chemistry employed is the *ab initio* HF/TZVP model, i.e. the Hartree-Fock theory and the TZVP basis set [16]. The details of this HF/TZVP model can be referred to our previous studies [3-4, 13]. The valence electron wave-functions of hexane produced using the HF/TZVP model are directly mapped into the momentum space [17]. The spherically averaged gamma-ray spectra of the valence electrons are then calculated using the equations in [3].

The structure of the normal hexane in three dimensional (3D) spaces is given in Fig.1 (a), together with the atom labelling and calculated atomic charges in brackets based on Mulliken population analysis [15, 18]. Table 1 gives the calculated molecular geometric properties of n-hexane and compared with available literature values. As n-hexane has the $C_{2h}$ point group symmetry so that only the unique geometric properties regarding the C-C and C-H bond lengths and bond angles are given in this table (other geometric parameters can be produced by point group symmetry). As shown in this table, the C-C and C-H bonds agree well with available results of Hunt and East [19]. Although small, the terminal C-C and C-H bonds, such as C(1)-C(6) and C(1)-H(9), are smaller than the centre bonds such as C(1)-C(4) and C(3)-H(17). The present study and Hunt and East [19] also agree that the C(3)-H(17) bond is the longest C-H bond of n-hexane. The C-C bond angles of the n-hexane are also in agreement with other literature results [20].

Fig.1 (b) reports the calculated total electron density of n-hexane mapped on the total molecular electrostatic potentila (ESP) with the positive (blue) and negative (red) potentials representing by the color scheme. It is suggested by a recent study that the attractive potential and the chemical environment of a molecule play important roles in the annihilation process [11]. The calculated total ESP is the electrostatic Coulomb interactions of the hexane system with a positron. The ESP in this figure can be employed as an indicator of the positrophilic sites in the annihilation processes which will be discussed later. It is well known in organic chemistry that all the C-H bonds of hydrocarbons are polar bonds although normal alkanes such as n-hexane do not possess a permanent dipole moment. As shown in Fig.1(b), the negative potentials (red) concentrate on the carbon atoms, whereas the positive potentials (blue) concentrate on the hydrogen atoms, which is in agreement with the Mulliken population charges indicated in Fig.1 (a). As indicated by Tachikawa et al in their recent study [21] that the positron is attached to the electronic negatively charged nitrogen atom of the C-N bond of the $CH_3CN$ molecule. In the case of hexane, the positron is likely to attach to the partially electronic negatively charged carbon atom of the C-H polar bond of n-hexane.

Fig1 (b) also indicated that the ESP distribution is not the same on all the carbon atoms nor on all the hydrogen atoms in the n-hexane, depending on the point group symmetry of hexane. For example, the negative potential (and charge, too) is more intensive at the terminal carbon atoms (i.e., C(1) and C(2)) and the C-C bond regions. As a result, these intensive negative potential regions are the more positron attractive regions. On the contrast, the partially positive potential regions such as vicinity of the hydrogen atoms are likely positron repulsive.

The probability of a positron to annihilate an electron from a molecular orbital $i$ of the target molecule is estimated as [7]

$$P_i = N \sum_j \left| C_{ij} \right|^2 , \qquad (1)$$

where $C_{ij}$ is the coefficient of the atomic orbital (basis set) $j$ to the molecular orbital $i$ and N is the normalization factor [7]. Equation (1) indicates that the probability of annihilation is target orbital dependent. As a result, the dominant orbitals which are determined by the largest $|C_{ij}|^2$ terms in Eq.(1) are the most probable orbitals in the annihilation process. In this study, we call the electrons in such the dominant orbitals of the target molecule are positrophilic electrons.

The positrophilic electrons of n-hexane are the valence electrons which dominate the Doppler-shift of the gamma-ray spectra in gas phase. In our HF/TZVP calculations, the ground electronic state ($X^1A_g$) of the normal hexane ($C_6H_{14}$) with a $C_{2h}$ point group symmetry has a closed shell configuration with 25 doubly occupied molecular orbitals which contains 19 valence orbitals. The calculated configuration ($X^1A_g$) of hexane contains six inner valence orbitals

$$(4a_g)^2(4b_u)^2(5a_g)^2(5b_u)^2(6a_g)^2(6b_u)^2, \qquad (2)$$

and thirteen outer valence orbitals

$$(1a_u)^2(1b_g)^2(7a_g)^2(7b_u)^2(2a_u)^2(8b_u)^2(8a_g)^2(2b_g)^2(9b_u)^2(3a_u)^2(9a_g)^2(3b_g)^2(10a_g)^2, (3)$$

where orbitals $4a_g$ and $10a_g$ are the inner most valence orbital and the highest occupied molecular orbital (HOMO), respectively. As shown in the quantum mechanical calculation of hexane that the inner valence orbitals (are dominated by carbon 2s orbitals) are more localised than the extensively delocalised outer valence molecular orbitals (dominated by the carbon 2p orbitals). Fig.2 gives the orbital distributions of the inner most orbital $4a_g$ and the HOMO, $10a_g$. As seen in this figure, the HOMO contains a number of nodal planes whereas the $4a_g$ orbital doesn't. As a result, the electrons in the HOMO possess higher (negative) energies than other valence electrons in the molecule.

With a sufficient energy positron-electron annihilation spectrum of a molecule is largely determined by the wavefunction (orbital) of a valence electron in the molecule.

A positron is accelerated in the vicinity of the target hexane by the molecule Coulomb attractive potential (red region in Fig.1 (b)) to annihilate a valence electron in the target. Valence electron distribution of the target hexane is determined by the wavefunctions (orbitals, such as those in Fig.2) which are dominated by the total ESP of hexane (balance of negative and positive components). As a result, the positron density closely relates to the negative potential of the target. Fig.3 shows positrophilic electrons $6b_u$, $6a_g$, and $4a_g$ of hexane, together with the superposition of the positrophilic electrons. It is noted that the positrophilic electrons of hexane all locate in the more concentrated inner valence space of n-hexane (Eq.(2)), in agreement with previous studies of alkanes that the electrons are annihilated underneath of the HOMO [7].

Fig.4 compares the calculated gamma-ray spectra of n-hexane in the positron-electron annihilation process with available experimental measurements of n-hexane [2, 12]. Note that the "×" and circle ("o") spectra are both from the same gas phase measurement of Surko et al [2]. The "×" spectrum is the original data points from the measurement without manipulation, whereas the "o"s are the two-Gaussian fit of the same measurement [2]. The red triangles ("Δ") are the earlier measurement of Kerr, Chuang and Hogg [12] of liquid n-hexane, which has been transformed into gamma-ray spectra from the measured momentum distributions [12] based on equations in [3]. The black solid spectrum is the calculated Doppler-shift in gas phase from the dominant positrophilic electrons of n-hexane, and the red dash spectrum represents the calculated Doppler-shift in liquid phase from all the valence electrons of n-hexane. Excellent agreement with the gas phase measurement (×) of Surko *et. al.* [2] and with the measurement in liquid phase (Δ) of Kerr *et. al.* [12] has been achieved in the present calculations. It is noted that in the same gas phase experiment of hexane [2], the Doppler-shift obtained from the actual experiment (×) is 2.45 keV whereas the two-Gaussian fit of the same measurement is given by 2.25 keV. That is, an experimental fitting/data handling related discrepancy in the gamma-ray measurement can be as large as ±0.20 keV (or ±8.2%).

The significant discrepancy in Doppler-shift between two measurements of n-hexane in gas phase [2] and liquid phase [12] indicates that the impact of the phase factor (i.e., gas phase and liquid phase) is considerably large impact on the measurements, apart from instrumental and technical issues. A significant difference between gas phase and liquid phase of the same substance is the intermolecular forces. In gas phase (vacuum), a molecule is almost free from intermolecular interactions. A positron most likely annihilates the positrophilic electrons of the free target molecule determined by the ESP of the molecular target, as the target moves free in the three dimensional space allowing the selectivity of such annihilation. These positrophilic electrons due to the non-spherically symmetric molecular ESP lead the incoming positron to be more selective than the spherical ESP of atoms in the annihilating process. The probability [7], $P_i$, of particular electrons annihilated by a positron which is determined by eq. (1) is not the same for all valence orbitals. As a result, a simple superposition (sum) contribution of all individual valence electrons to the gamma-ray spectra implies an assumption that the probability of every valance electron in the target molecule is the same. This assumption underestimates the positrophilic electrons or over estimates the electrons with only minor roles in the annihilation as previously seen in small molecules [13]. For example, the Doppler-shift ($\Delta\varepsilon$) of the positrophilic electrons of $O_2$ ($2\sigma_g$) is 2.87 keV ($\Delta\varepsilon_{exp}$ is 2.73 keV), whereas such superposition of all valence electrons ($\Delta\varepsilon$) of $O_2$ is 4.00 keV and all electrons (core + valence) ($\Delta\varepsilon$) is 4.22 keV [13].

In liquid phase, however, the hexane molecules aggregate together by intermolecular forces. The hexane molecules cannot move freely in the liquid phase, rather, the molecules are confined in the vicinity of ether other in liquid phase. The neutralization of the positrophilic and electrophilic areas of the target in liquid eliminates the selectivity of the positrophilic valence electrons, so that all valence electrons of the target need to be included in the spectra.

In summary, the present study calculates gamma-ray spectra of hexane in gas phase and liquid phase, respectively, using the positrophilic electron model and all-valence electron model. The results are in excellent agreement with earlier measurements of the same molecule in gas phase [2] as well as in liquid phase [12]. Such the excellent agreement with measurements confirms that the gamma-ray spectra of hexane are indeed dominated by the positrophilic electrons in gas phase, supported by a number of previous studies [3, 5, 7, 13]. The present study further indicates that positrophilic electrons of a target molecule are determined by the electrostatic potential (ESP) of the target molecule and the positron. While in gas phase the positrophilic electrons dominate contributions to the gamma-ray spectra, the intermolecular interactions in hexane liquid phase eliminate the selectivity of the electrons so that all valence electrons in liquid contribute to the annihilation process. To our knowledge, it is the first time that the present study accurately calculates the Doppler-shift of gamma-ray spectra of hexane in gas and liquid conditions using *ab initio* quantum mechanical methods, supported by experimental measurements.

This project is supported by the Australia Research Council (ARC) under the Discovery Project (DP) Scheme. National Computational Infrastructure (NCI) at the Australia National University (ANU) under the Merit Allocation Scheme (MAS) and the Swinburne University of Technology gSTAR supercomputer should be acknowledged. The authors also acknowledge Prof. C. M. Surko for providing the gas phase experimental data of hexane.

Table 1 Geometric parameters of n-hexane (gas phase) obtained using the HF/TZVP model*.

| Hexane(C$_{2h}$) | | This work | Literature values |
|---|---|---|---|
| Bond lengths (Å) | C(1)-C(6) | 1.5337 | 1.5246[19] |
| | C(3)-C(6) | 1.5358 | 1.5256[19] |
| | C(3)-C(4) | 1.5355 | 1.5254[19] |
| | C(1)-H(8) | 1.1037 | - |
| | C(1)-H(9) | 1.1048 | 1.090[19] |
| | C(6)-H(13) | 1.1075 | 1.090[19] |
| | C(3)-H(17) | 1.1087 | 1.094[19] |
| Bond angles (°) | C(2)-C(5)-C(4) | 113.163 | 113.6[20] |
| | C(5)-C(4)-C(3) | 113.607 | 114.1[20] |

*The atomic labeling scheme of n-hexane refers to Figure 1(a).

Figure captions

Figure 1: (a) Bird-view of the structure and atomic labbeling scheme of n-hexane
($C_6H_{14}$). Mulliken population charges are also given on each atoms. Note
that due to the $C_{2h}$ point group symmetry, the atomic labbels are based on
the unique atoms. (b) The total electrostatic potential (ESP) of hexane is
mapped on the optimised ball-stick structure of n-hecane. The colour bar
represents the values of the ESP (red: negative and blue: positive).

Figure 2: The orbital distributions of the inner most orbital *$4a_g$* and the HOMO, *$10a_g$*.

Figure 3: (a) The contour color-filled map of the positrophilic orbitals of $4a_g$, $6a_g$, $6b_u$
and $6b_u+6a_g+4a_g$. (b) 3D-surface of negative potential (orange) with the
electron density of the positrophilic electrons "$6b_u+6a_g+4a_g$" (cyan).

Figure 4: Gamma-ray spectra of n-hexane molecule in positron-electron annihilation
process compared with two experimental measurements. The numbers in
brackets are the Full Widths at Half Maximum (FWHM, i.e. the Doppler
shift) in keV.

× is the measurement for the low-pressure gas phase provided by Surko[2].

Δ is the measurement of organic liquid hexane of Kerr et al [12].

o is the two-Gaussian fitted experimental spectra from the same experiment of
Surko et al [2].

—— is the caluclated gamma-ray spectra of n-hexane in gas phase obtained from
dominant positrophilic electrons of hexane and

- - - is calculated gamma-ray spectra of n-hexane in liquid phase obtained
from supperposition of all valence electrons of hexane. All spectra are
normalized to unity at zero.

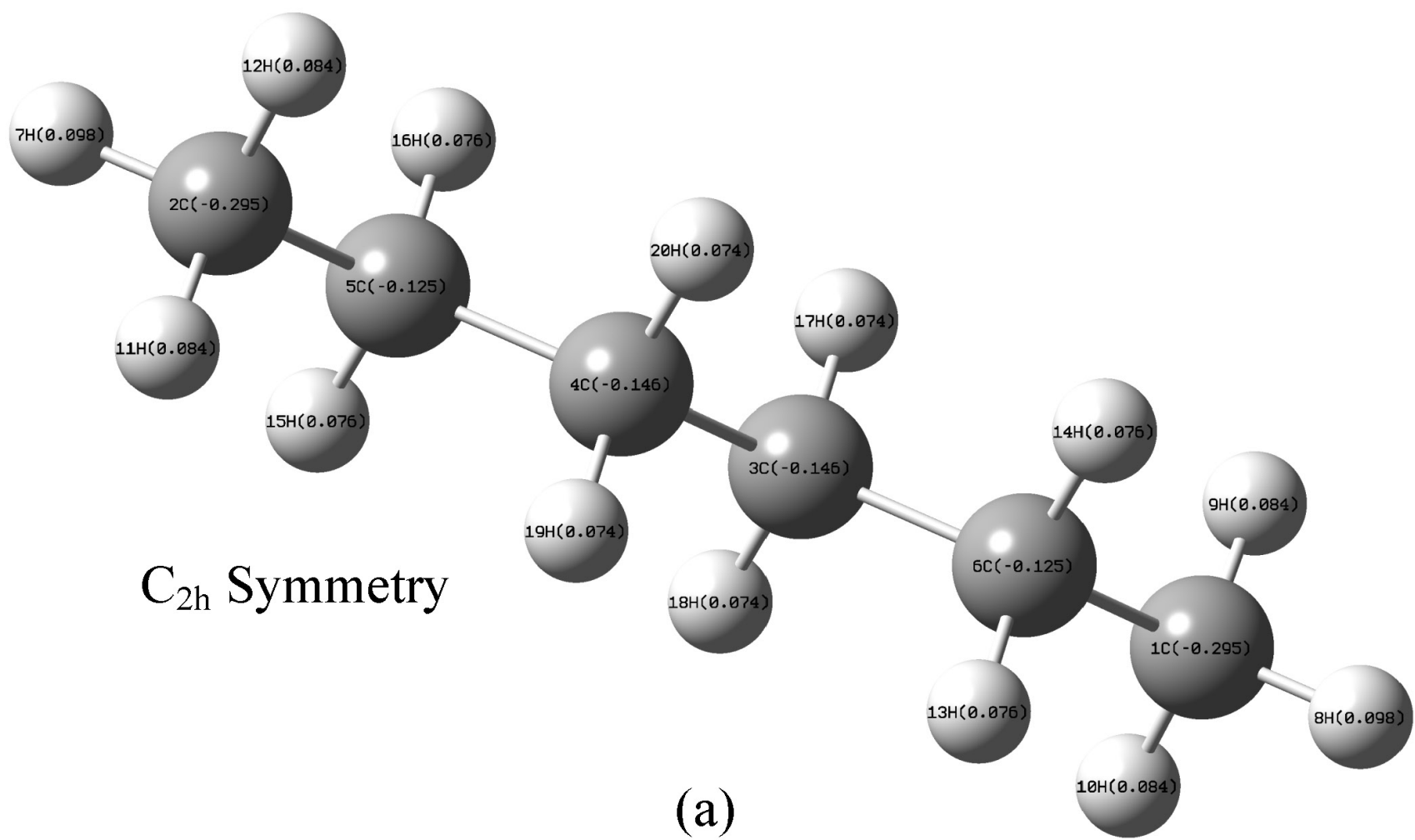

C$_{2h}$ Symmetry

(a)

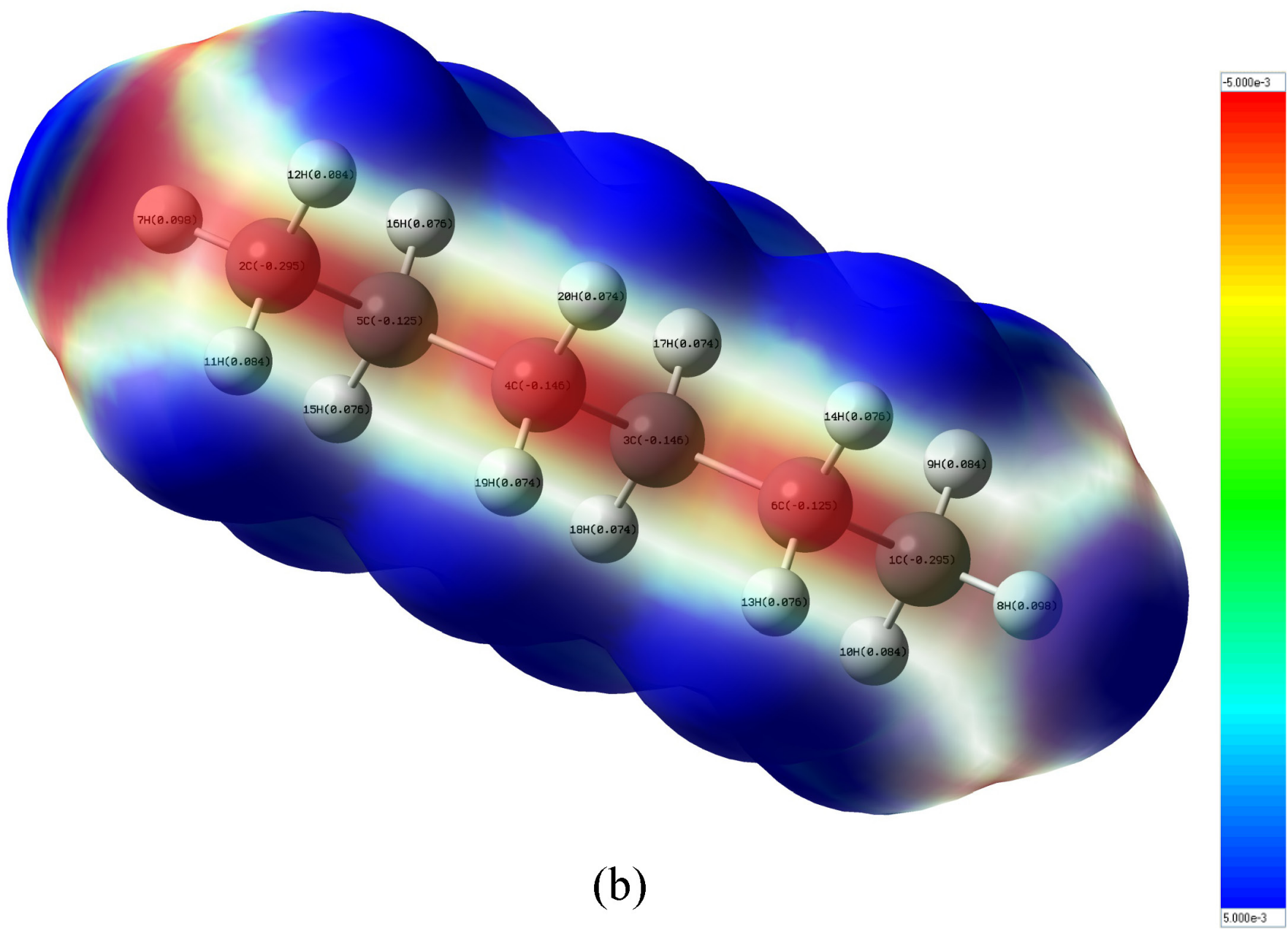

(b)

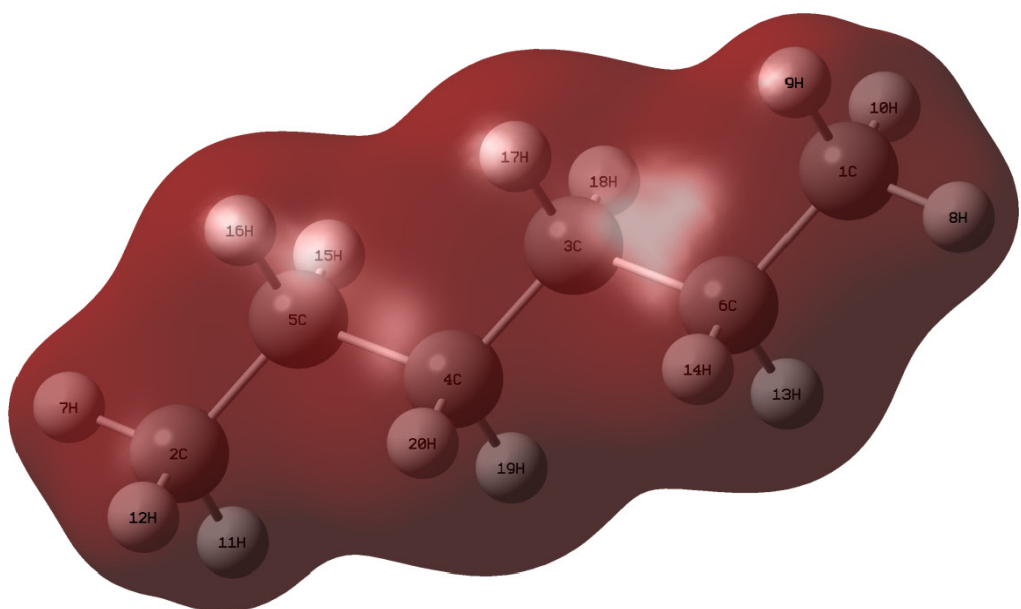

$4a_g$

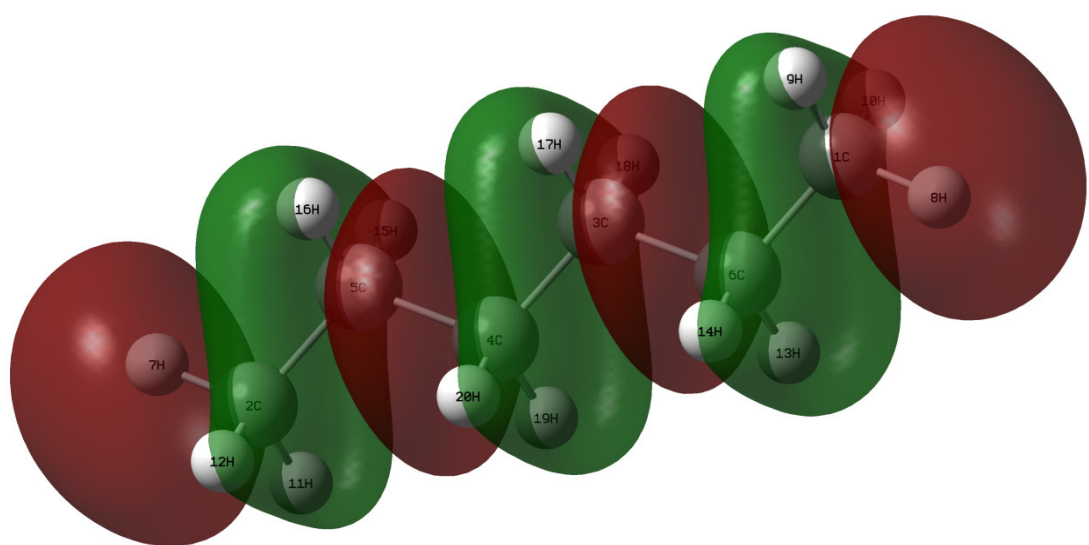

$10a_g$

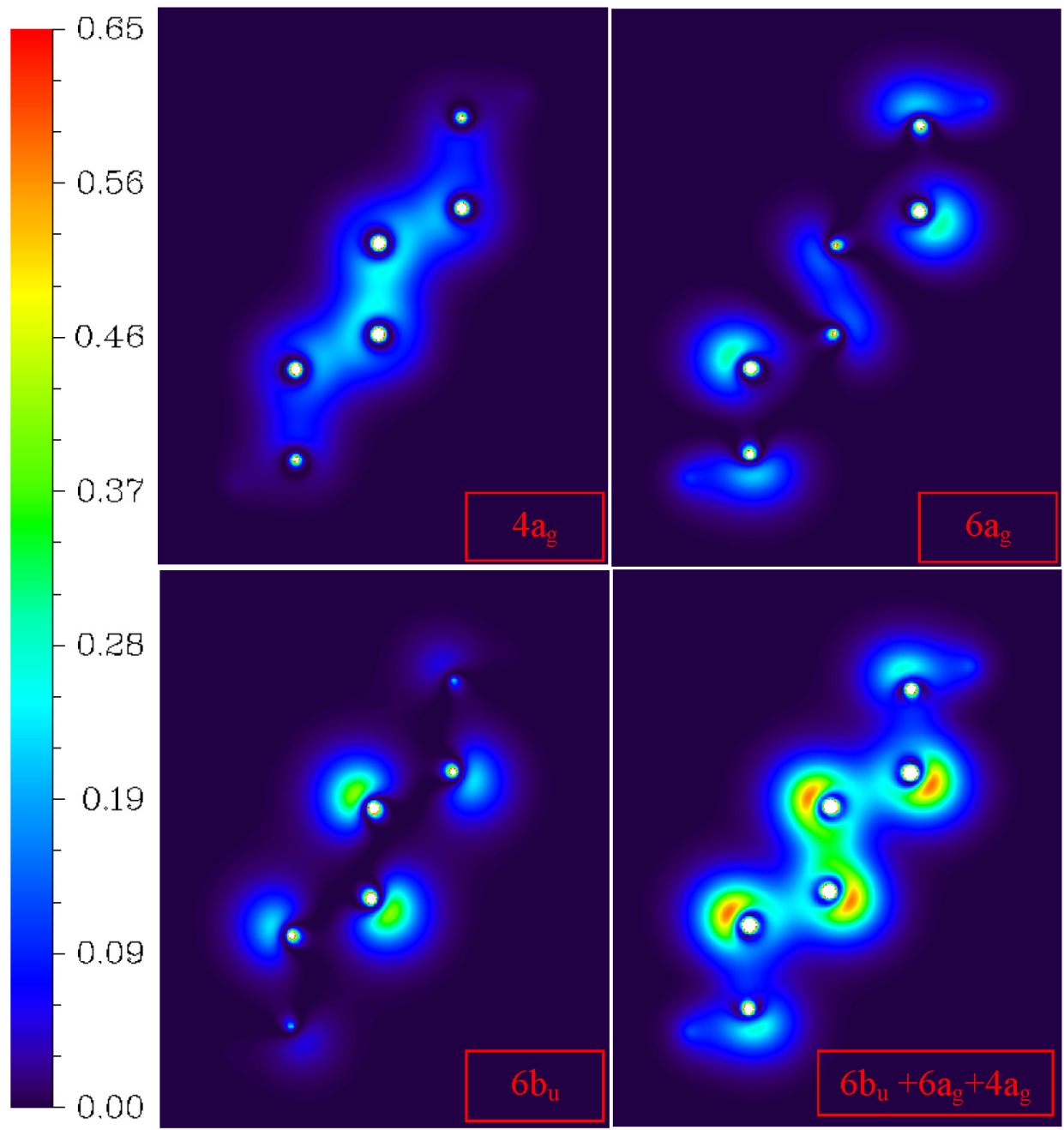

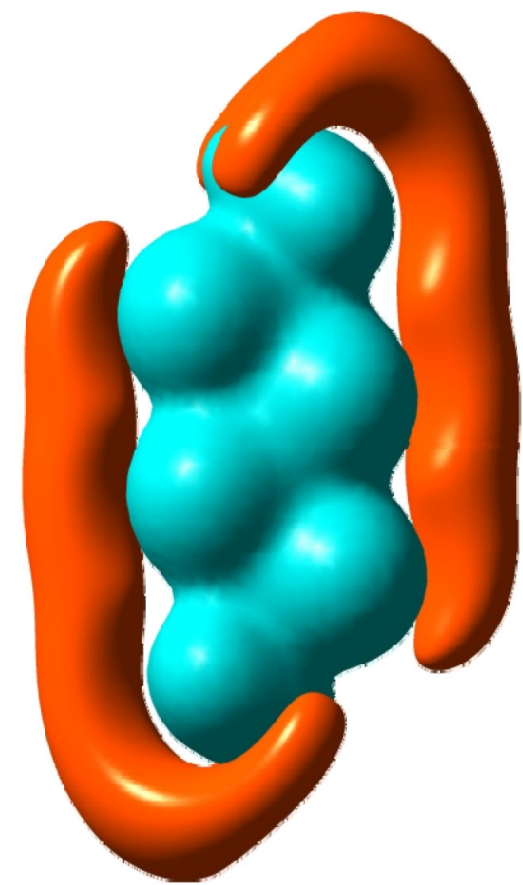

(a)

(b)

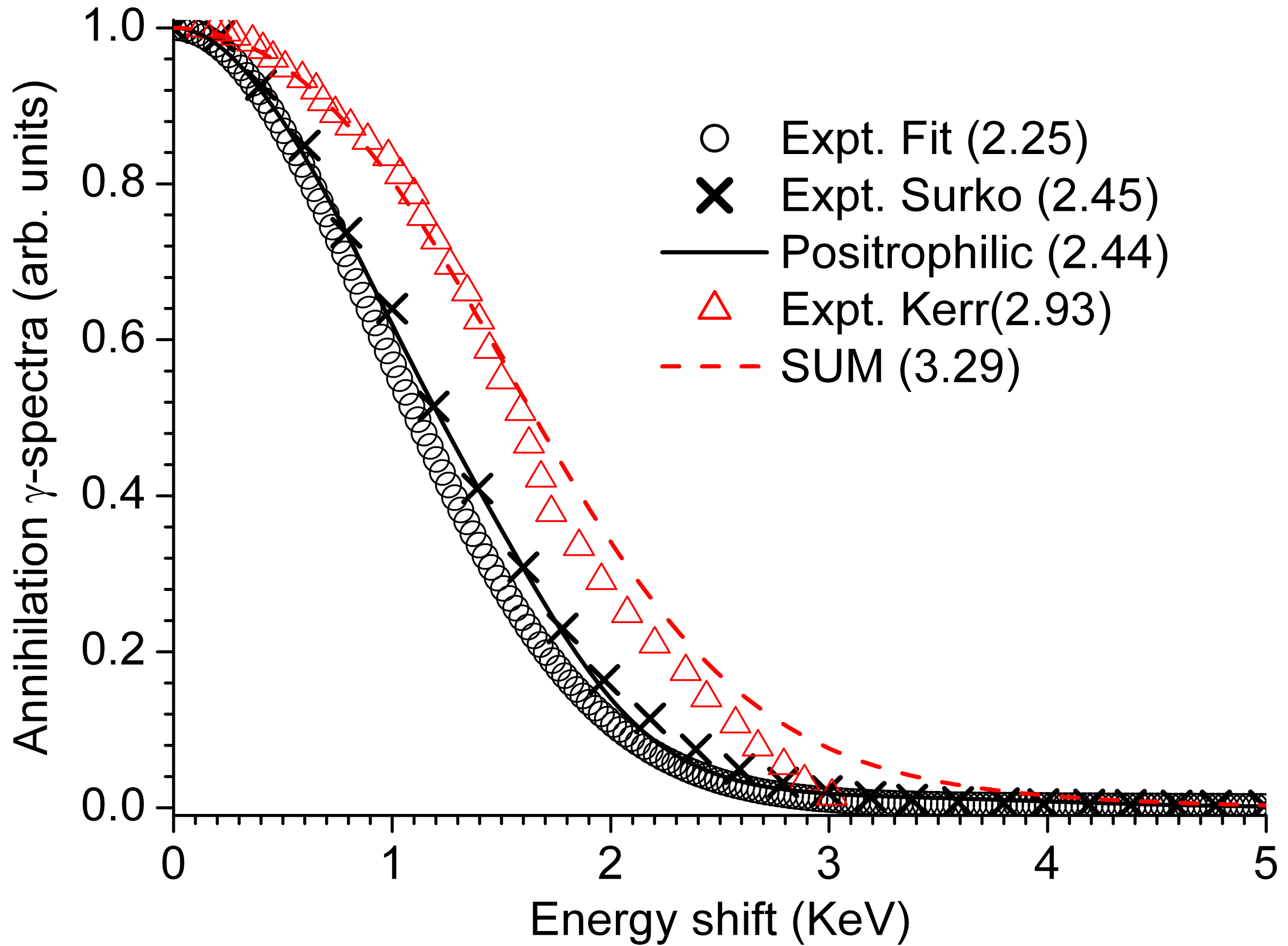